\documentclass[conference]{IEEEtran}
\IEEEoverridecommandlockouts

\usepackage{cite}
\usepackage{amsmath,amssymb,amsfonts}
\usepackage{algorithmic}
\usepackage{graphicx}
\usepackage{textcomp}
\def\BibTeX{{\rm B\kern-.05em{\sc i\kern-.025em b}\kern-.08em
    T\kern-.1667em\lower.7ex\hbox{E}\kern-.125emX}}
\usepackage{multirow}
\usepackage{makecell}
\usepackage[dvipsnames]{xcolor}

\usepackage[draft]{minted}

\usepackage{listings}
\usepackage{xcolor}
\usepackage{hyperref}

\usepackage{pgfplots}
\pgfplotsset{compat=1.17}
\pgfplotsset{compat = newest}

\definecolor{my_dark_violet}{RGB}{90, 24, 154}
\definecolor{my_light_violet}{RGB}{199, 125, 255}
\definecolor{my_dark_green}{RGB}{56, 102, 65}
\definecolor{my_light_green}{RGB}{167, 201, 87}
\definecolor{my_light_blue}{RGB}{0, 180, 216}
\definecolor{my_dark_blue}{RGB}{2, 62, 138}
\definecolor{my_dark_red}{RGB}{208, 0, 0}
\definecolor{my_light_red}{RGB}{244, 140, 6}
\definecolor{my_dark_teal}{RGB}{17, 157, 164}
\definecolor{my_light_teal}{RGB}{174, 236, 239}
\definecolor{my_gray}{RGB}{141, 153, 174}
\definecolor{my_yellow}{RGB}{255, 234, 0}
\definecolor{my_pink}{RGB}{255, 179, 198}

\lstdefinelanguage{Verilog}{
  morekeywords={module, endmodule, input, output, wire, reg, assign, always, if, else, case, endcase, begin, end},
  sensitive=true,
  morecomment=[l]{//},
  morecomment=[s]{/*}{*/},
  morestring=[b]",
}

\begin{document}

\title{VeriMind: Agentic LLM for Automated Verilog Generation with a Novel Evaluation Metric\\
}

\author{\IEEEauthorblockN{1\textsuperscript{st} Bardia Nadimi}
\IEEEauthorblockA{\textit{Dept. of Computer Science \& Eng.} \\
\textit{University of South Florida}\\
Tampa, Florida, United States \\
bnadimi@usf.edu}
\and
\IEEEauthorblockN{2\textsuperscript{nd} Ghali Omar Boutaib}
\IEEEauthorblockA{\textit{Dept. of Computer Science \& Eng.} \\
\textit{University of South Florida}\\
Tampa, Florida, United States \\
ghaliomar@usf.edu}
\and
\IEEEauthorblockN{3\textsuperscript{rd} Hao Zheng}
\IEEEauthorblockA{\textit{Dept. of Computer Science \& Eng.} \\
\textit{University of South Florida}\\
Tampa, Florida, United States \\
haozheng@usf.edu}

}

\maketitle

\begin{abstract}

Designing Verilog modules requires meticulous attention to correctness, efficiency, and adherence to design specifications. 
However, manually writing Verilog code remains a complex and time-consuming task that demands both expert knowledge and iterative refinement. 
Leveraging recent advancements in large language models (LLMs) and their structured text generation capabilities, we propose VeriMind, an agentic LLM framework for Verilog code generation that significantly automates and optimizes the synthesis process. 
Unlike traditional LLM-based code generators, VeriMind employs a structured reasoning approach: given a user-provided prompt describing design requirements, the system first formulates a detailed train of thought before the final Verilog code is generated. 
This multi-step methodology enhances interpretability, accuracy, and adaptability in hardware design. 
In addition, we introduce a novel evaluation metric—pass@ARC—which combines the conventional pass@k measure with Average Refinement Cycles (ARC) to capture both success rate and the efficiency of iterative refinement. 
Experimental results on diverse hardware design tasks demonstrated that our approach achieved up to $8.3\%$ improvement on pass@k metric and $8.1\%$ on pass@ARC metric. 
These findings underscore the transformative potential of agentic LLMs in automated hardware design, RTL development, and digital system synthesis.

\end{abstract}


\vspace*{-5pt}
\section{Introduction}
\label{sec:introduction}

With the introduction of attention-based models and transformers \cite{attentionIsAllYouNeed}, the field of natural language processing (NLP) underwent a significant transformation. 
In recent years, transformers have become a fundamental component in the architecture of most AI models. 
Various models, such as Generative Pre-trained Transformers (GPT) \cite{radford2018GPT}, Bidirectional Encoder Representations from Transformers (BERT) \cite{BERT}, Language Model for Dialogue Applications (LaMDA) \cite{LaMDA}, and Vision Transformers (ViT) \cite{ViT}, have utilized transformer architectures to push the boundaries of performance across natural language processing and computer vision tasks.

The success of transformers stems from their ability to effectively model long-range dependencies, enabling them to capture contextual relationships in textual, visual, and multimodal data with unprecedented accuracy. 
This has led to significant advancements in diverse applications, including text generation, machine translation, code synthesis, and image understanding. 
Despite these achievements, applying transformers to specialized domains such as hardware design and Verilog code generation remains relatively unexplored. 
The complexity of hardware description languages (HDLs), the need for logical correctness, and the structured nature of Verilog pose unique challenges that traditional AI-driven code generation techniques struggle to address.
The utilization of large language models (LLMs) for hardware code generation is driven by the goal of developing an advanced tool that simplifies the hardware modeling process for designers \cite{MEV-LLM, MaskedLM, spicepilot, SA-DS}. 
Furthermore, automating hardware code generation using LLMs plays a crucial role in minimizing human error. 
Due to the inherently complex and highly technical nature of hardware design, manual coding is susceptible to inaccuracies and inconsistencies \cite{Hardfails}.
Integrating these models into hardware development workflows reduces the inherent complexities of traditional design methods. 
Leveraging LLM capabilities enables a more intuitive and efficient approach to hardware modeling, thereby enhancing automation and design optimization.
In addition, ensuring the security of LLM-generated code is paramount \cite{security1, security2}; rigorous code validation and adversarial testing must be implemented to safeguard against vulnerabilities and unauthorized modifications.

As AI-driven automation continues to expand, leveraging agentic AI for Verilog code generation presents a promising opportunity to enhance productivity, reduce human effort, and improve design efficiency in hardware development. 
In this work, we introduce VeriMind, an agentic AI approach for Verilog code generation that leverages the reasoning and iterative refinement capabilities of LLMs. 
Unlike conventional AI-driven code generation, which passively produces code based on a single prompt, our method employs a structured multi-step process to enhance the accuracy, interpretability, and adaptability of the generated hardware descriptions. 
By integrating cognitive reasoning and planning, our framework systematically decomposes the Verilog generation task, enabling the AI to outline a logical design process before synthesizing the final code.
Furthermore, to facilitate a more comprehensive evaluation of approaches that employ refinement loops, we introduce a novel metric termed pass@ARC (Average Refinement Cycles). 
This metric integrates the traditional pass rate with the number of refinement iterations required to generate a correct code. 
By penalizing models that require more cycles, pass@ARC provides a more nuanced assessment of both accuracy and efficiency, making it an invaluable tool for comparing and optimizing iterative code generation architectures.

Our method adopts an interactive, self-improving workflow. 
Upon receiving a design specification from the user, the proposed architecture iteratively refines the generated Verilog code—ensuring both syntactic and functional correctness—with dedicated agents managing specific subtasks. 
A detailed description of the architecture and its individual agents is provided in Section \ref{sec:methodology}.

\begin{figure}[t]
    \centering
    \includegraphics[width=\columnwidth]{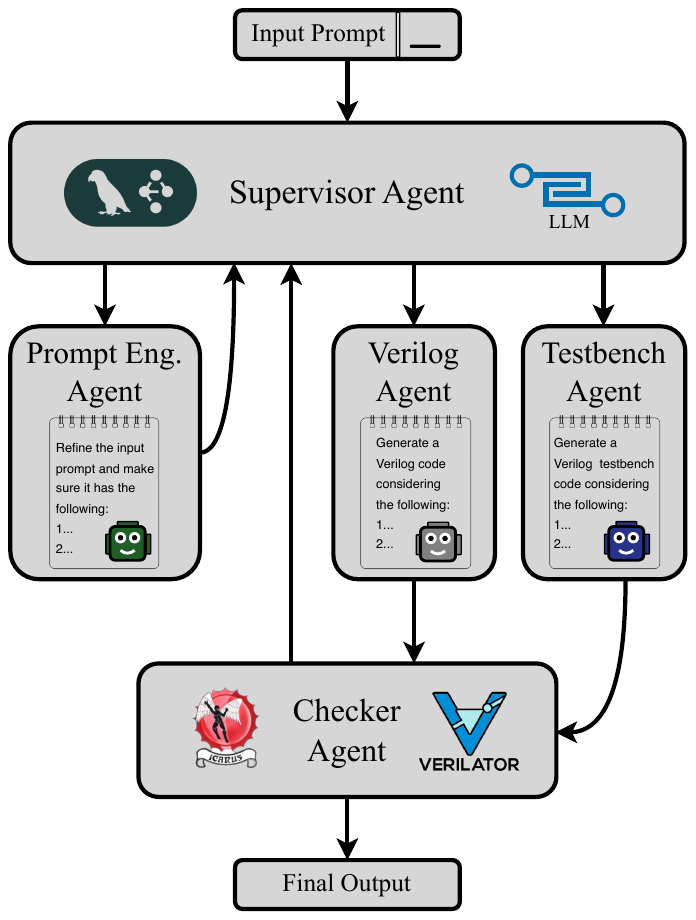}
    \caption{Proposed Overall Architecture.}
    \label{fig:overallArchitecture}
    \vspace*{-15pt}
\end{figure}

To evaluate the effectiveness of our method, we apply it to a diverse set of Verilog design tasks, ranging from simple combinational circuits to complex sequential logic. 
Our results demonstrate that agentic AI significantly improves the quality, correctness, and adaptability of generated Verilog code compared to traditional LLM-based code generation. 
This work highlights the potential of integrating structured reasoning and interactive validation into AI-driven hardware development, paving the way for more autonomous and reliable HDL design automation.
The overall architecture of the proposed approach is illustrated in Fig.\ref{fig:overallArchitecture}. The \textbf{key contributions} of this paper are summarized as follows:
\begin{itemize}
    \item Agentic AI Framework for Verilog Code Generation: Introduces a structured, reasoning-driven approach that enhances interpretability, correctness, and adaptability in hardware description language (HDL) generation.
    \item Interactive and Iterative Refinement Process: Incorporates intermediate validation steps where the AI outlines its reasoning before generating Verilog code, allowing user feedback to guide improvements.
    \item Enhanced Code Quality and Design Consistency: Evaluates the proposed approach across diverse Verilog design tasks, demonstrating improved syntactic correctness, logical coherence, and alignment with design specifications compared to conventional LLM-based generation.
    \item We introduce a novel evaluation metric, pass@ARC, specifically designed to provide a fair and comprehensive assessment of architectures that employ iterative refinement loops.
\end{itemize}

The remainder of this paper is structured as follows: Section \ref{sec:relatedWorksAndBackgrounds} presents an overview of related work and background information. 
Section \ref{sec:methodology} details the proposed architecture. 
Section \ref{sec:evaluationAndDiscussion} discusses the evaluation results and analysis. 
Finally, Section \ref{sec:conclusionAndFutureWorks} outlines the conclusions and potential directions for future research.
\vspace*{-4pt}
\section{Related Works and Backgrounds}
\label{sec:relatedWorksAndBackgrounds}

High-quality Verilog datasets are crucial yet challenging for effective hardware code generation with LLMs. 
Nadimi et al.\cite{pyranet} addressed this by developing PyraNet, an open-source, hierarchical dataset that enhances the syntactic and functional accuracy of generated Verilog code. 
By using quality-based tiers and curriculum learning, models fine-tuned with PyraNet improved performance by up to $32.6\%$ over CodeLlama-7B and $16.7\%$ over state-of-the-art models on VerilogEval.

Advancements in Verilog dataset augmentation and LLM fine-tuning include approaches such as CodeV \cite{CodeV} and MG-Verilog \cite{mgverilog}. 
CodeV employs multi-level summarization on a dataset of 165,000 Verilog modules enriched with GPT-3.5 annotations, achieving state-of-the-art results on VerilogEval and RTLLM—surpassing GPT-4 by $22.1\%$ on VerilogEval. 
Similarly, MG-Verilog provides a multi-granular dataset of over 11,000 samples, enhancing LLM generalization and improving code synthesis accuracy.

Beyond dataset-driven improvements, several studies have optimized Verilog generation through fine-tuning, retrieval mechanisms, and structured prompting. 
BetterV \cite{BetterV} uses instruct-tuning with generative discriminators and Verilog-C program alignment to enhance key EDA processes. 
Meanwhile, AutoVCoder \cite{AutoVCoder} leverages retrieval-augmented generation with a two-phase fine-tuning strategy to better align general-purpose LLMs with RTL constraints, achieving a $3.4\%$ improvement in functional correctness on the RTLLM benchmark. 
This demonstrates the significant impact of retrieval-enhanced architectures on RTL automation.

Another important line of research focuses on self-reflective learning and compiler-based error correction. OriGen \cite{origen} introduces a self-correcting framework that combines code-to-code augmentation with compiler feedback loops for error detection and refinement. OriGen trains two specialized Low-Rank Adaptation (LoRA) models—one for initial Verilog generation and another for syntax correction—resulting in a 12.8\% performance boost over existing open-source models and a 19.9\% improvement in syntactic accuracy compared to GPT-4 Turbo. Incorporating iterative self-reflection, OriGen greatly improves the reliability of LLM-generated Verilog.

To tackle challenges related to hierarchical Verilog generation, Tang et al. \cite{hivegen} proposed HiVeGen, a structured framework that decomposes HDL synthesis into hierarchical submodules. Traditional LLMs often generate monolithic, hallucination-prone HDL blocks, particularly for complex designs like domain-specific accelerators. HiVeGen mitigates this issue by introducing automated design space exploration, structured code reuse, and human-computer interaction mechanisms to reduce debugging overhead. Empirical evaluations show that HiVeGen enhances the quality and scalability of AI-generated HDL, making it a promising approach for modular chip design automation.

Structured prompting techniques have been explored to enhance Verilog synthesis. 
ROME \cite{rome} employs hierarchical prompting to decompose large hardware modules into organized submodules and integrates an eight-stage automation pipeline that combines human-guided and autonomous refinement. 
Evaluations show that fine-tuned LLMs using ROME outperform traditional flat prompting in generating complex HDL, reducing both development time and computational overhead. 
Notably, ROME produced the first LLM-designed processor without human intervention, underscoring its revolutionary potential in AI-assisted chip design.

Multi-agent frameworks have proven beneficial for RTL design by enhancing efficiency, interpretability, and performance. 
RTLSquad \cite{RTLsquad} introduces a collaborative system that divides the RTL workflow into three phases—exploration, implementation, and verification—each managed by specialized agent squads to ensure seamless communication and transparent decision-making. 
Experimental results demonstrate that RTLSquad not only produces functionally correct RTL code but also improves power, performance, and area metrics, underscoring its value in scalable AI-assisted hardware development.
\vspace*{-6pt}
\section{Methodology}
\label{sec:methodology}
This section begins with an overview of the complete agentic architecture, providing a high-level understanding of its structure and functionality. 
Subsequently, each agent is examined in detail, outlining its specific role and operational mechanisms within the framework.
\vspace*{-4pt}

\subsection{Agentic Framework: Structure and Workflow}
\label{subsec:overallArchitecture}

Our proposed architecture as shown in Fig.~\ref{fig:overallArchitecture} comprises five interconnected agents—Supervisor, Prompt Engineer, Verilog Code Generator, Verilog Testbench Generator, and Checker, that collaborate to generate verified Verilog designs. 
The Supervisor Agent orchestrates the workflow by initially sending the user’s prompt to the Prompt Engineer Agent for refinement. 
The refined prompt is then dispatched to both the Verilog Code Generator and Testbench Generator Agents to produce the corresponding outputs, which are subsequently validated by the Checker Agent through compilation and simulation. 
Should errors arise, the Checker Agent triggers a re-evaluation cycle via the Supervisor Agent until the final, error-free Verilog code is produced. 
This iterative approach ensures high accuracy, functionality, and reliability in the generated designs, with further details on each agent provided in the following sections.
\vspace*{-4pt}

\subsection{Supervisor Agent}
\vspace*{-2pt}

The Supervisor Agent is the central coordinator in our multi-agent architecture for Verilog code generation, orchestrating the entire process from prompt refinement to final verification. 
This agent employs a lightweight LLM solely to manage inter-agent communications, including selecting appropriate agents at various stages and modifying prompts as necessary. 
In our experiments, we employed the gpt-4o-mini model for this agent.
The process begins by receiving the initial design specification from the user and immediately dispatches this prompt to the Prompt Engineer Agent. 
This agent refines and optimizes the prompt, ensuring that it is well-structured for subsequent processing. 
Once the refined prompt is obtained, the Supervisor Agent simultaneously sends it to both the Verilog Code Generator Agent and the Verilog Testbench Generator Agent, enabling parallel generation of the Verilog module and its corresponding testbench.

After these agents complete their tasks, the generated outputs are forwarded to the Checker Agent, which validates the code by performing compilation and simulation tests. 
If the Checker Agent detects any errors—such as compilation failures, simulation mismatches, or logical inconsistencies—it promptly notifies the Supervisor Agent. 
The Supervisor then evaluates the reported issues and decides whether to re-send the prompt for further refinement, request corrections from the Verilog Code Generator, or ask the Testbench Generator to adjust its output. 
This iterative loop continues until the Checker Agent confirms that the generated Verilog design is both syntactically correct and functionally robust.

Moreover, the Supervisor Agent employs a structured message passing mechanism that ensures effective inter-agent communication and maintains a detailed state tracking system. 
This system monitors each agent’s progress and allows the Supervisor to dynamically adjust execution priorities in the event of delays or unexpected outcomes, thereby optimizing overall workflow efficiency.

Overall, the Supervisor Agent plays a critical role by seamlessly integrating specialized tasks—refinement, code generation, testbench generation, and validation—into a cohesive and adaptive process. 
This structured and iterative approach not only enhances the quality and accuracy of the generated Verilog modules but also significantly reduces manual intervention and error rates in hardware design.
\vspace*{-4pt}

\subsection{Prompt Engineer Agent}

The Prompt Engineer Agent refines the input prompt to ensure that the generated Verilog code is structurally sound, functionally relevant, and aligned with the user's intent. 
By modifying and reformatting the input, it increases the likelihood of producing high-quality code that meets design specifications. 
Like the Supervisor Agent, this agent uses a compact LLM—in our evaluation, we employed the gpt-4o-mini model for this role.

Upon receiving a raw prompt from the Supervisor Agent, the Prompt Engineer Agent enhances clarity, structure, and specificity. 
This process involves rewriting ambiguous specifications, expanding under-specified prompts with necessary design constraints (e.g., clocking details, functional properties), reformatting inputs into structured templates, and removing redundant information. 
For example, a vague request such as “Generate a Verilog module for a 4-bit counter” may be transformed into “Generate a Verilog module for a 4-bit synchronous up-counter with a clock input (clk), an asynchronous reset (rst), and an enable signal (en), where the output (count) is a 4-bit register.”
 
Additionally, the agent tailors its refinements to the task at hand—whether it involves combinational logic, sequential circuits, or finite state machines—by ensuring that the prompt includes all necessary details (e.g., Boolean expressions for combinational circuits or explicit state transitions for FSMs). 
If validation errors occur later, the Checker Agent informs the Supervisor Agent, prompting further modifications until an effective prompt is achieved.
\vspace*{-5pt}

\subsection{Verilog Code Generator Agent}
\label{subsec:VerilogAgent}
The Verilog Code Generator Agent uses a refined prompt from the Supervisor Agent and a specialized LLM for HDL generation to produce Verilog modules that are both structurally sound and functionally accurate. 
It is essential to the multi-agent framework, ensuring adherence to digital design best practices while maintaining efficiency.
This agent employs a fine-tuned LLM to manage the sophisticated task of Verilog code generation. 
For comparative analysis, we evaluated our architecture using a range of LLMs in this role, including both fine-tuned models and commercial alternatives.

Once the Prompt Engineer Agent refines the input request, the Supervisor Agent dispatches the structured prompt to the Verilog Code Generator Agent, initiating the code synthesis process. 
The Verilog Code Generator Agent is designed to handle a diverse range of digital designs, ensuring adaptability across different hardware modules:

\begin{itemize}
    \item \textbf{Combinational Circuits:} Generates Verilog for logic gates, multiplexers, arithmetic operations, and other stateless logic.
    \item \textbf{Sequential Circuits:} Implements flip-flops, registers, counters, and clocked state transitions.
    \item \textbf{Finite State Machines (FSMs):} Constructs Mealy and Moore state machines, including state encoding and transition logic.
    \item \textbf{Parameterized Modules:} Creates reusable Verilog components with adjustable parameters (generate statements, parameter values).
\end{itemize}

For example, when generating an FSM for a traffic light controller, the agent ensures that state transitions are clearly defined and that the design follows a systematic approach; a representative sample is provided in Appendix A.
\vspace*{-5pt}

\subsection{Testbench Generator Agent}
\vspace*{-2pt}
The Verilog Testbench Generator Agent is responsible for creating a corresponding testbench to verify the functionality of the generated Verilog module. 
By simulating various input conditions and monitoring the expected outputs, this agent ensures that the generated hardware design behaves as intended.
Mirroring the approach used by the Verilog Code Generator Agent, this agent utilizes a fine-tuned model for testbench generation. 
For comparative analysis, we evaluated our architecture with a variety of LLMs in this role, including both fine-tuned models and commercial alternatives.

Upon receiving the refined prompt, the Supervisor Agent instructs the Verilog Testbench Generator Agent to create a structured testbench. 
This process involves: 1) defining stimuli and input signals to drive the module under test (MUT), 2) instantiating the Verilog module within the testbench framework, 3) creating a simulation sequence, including clock generation, reset initialization, and test scenarios, and 4) using assertions or \$monitor statements to validate expected outputs.
For instance, the testbench generated for the traffic light controller described in Section \ref{subsec:VerilogAgent} is presented in Appendix B.

If the Checker Agent detects issues during compilation or simulation, the Supervisor Agent instructs the Verilog Testbench Generator Agent to modify the testbench accordingly. 
This iterative process ensures that: 1) the testbench correctly stimulates edge cases and corner conditions, 2) the verification covers all functional aspects of the Verilog module, and 3) the generated testbench is efficient, concise, and reusable for validation purposes.
\vspace*{-4pt}

\subsection{Checker Agent}
The Checker Agent ensures the accuracy of the generated Verilog module and its corresponding testbench by conducting compilation and simulation tests. 
Serving as a verification layer, it guarantees that the output complies with Verilog syntax rules and meets functional expectations. 
This agent leverages compilation and simulation tools, including Icarus Verilog \cite{IcarusVerilog} and Verilator \cite{verilator}, to systematically validate the correctness of the generated HDL code.

Upon receiving the Verilog module and testbench, the Checker Agent conducts a structured validation process to ensure the correct functionality of the design. 
It first compiles the Verilog module and testbench, checking for syntax errors, missing dependencies, or structural inconsistencies. 
Once compilation is successful, the agent performs a simulation to verify functional correctness, ensuring that the module operates as expected across different test scenarios. 
During the simulation, it monitors signal behavior and assertions using \emph{\$display}, or assertion-based verification methods to track discrepancies in output.

If the Checker Agent encounters issues—such as compilation errors, logical inconsistencies, or incorrect simulation results—it notifies the Supervisor Agent with such information, prompting a refinement loop where the Verilog Code Generator Agent or Verilog Testbench Generator Agent modifies the output to resolve the detected errors. 
This iterative validation process ensures that the final Verilog module is both syntactically correct and functionally reliable before being delivered to the user.

\section{Evaluation and Discussion}
\label{sec:evaluationAndDiscussion}
To evaluate our agentic AI framework for Verilog code generation, we used the VerilogEval benchmark suite \cite{verilogEval}. 
Recognizing that refinement loops require a nuanced evaluation, we introduced the pass@ARC metric—which combines pass@k with Average Refinement Cycles (ARC) (details in subsequent subsections). 
These benchmarks standardize the assessment of syntactic correctness, functional accuracy, and iterative efficiency.
\vspace*{-5pt}

\subsection{Experimental Setup}
\textbf{Benchmarks:} VerilogEval benchmark consists of two parts: 1) VerilogEval Machine and 2) VerilogEval Human with 143 and 156 problems respectively. 
They all sourced from the HDLbits platform\footnote{\href{https://hdlbits.01xz.net/wiki/Problem_sets}{https://hdlbits.01xz.net/wiki/Problem\_sets}}, covering a wide range of Verilog codes from combinational circuits to complex sequential circuits.
In addition, for the pass@ARC metric, we extend the VerilogEval platform by incorporating the average number of refinement cycles into the evaluation framework.

\textbf{Experimental setup:} For the comparison purposes, we implemented the proposed design using multiple baseline LLMs including GPT-4o-mini, Claude3.5-Sonnet, CodeLlama-70b-Instruct, DeepSeek-Coder-V2-Instruct and PyraNet. 
We compared the results with other SOTA works including PyraNet \cite{pyranet}, Origen \cite{origen}, AutoVCoder \cite{AutoVCoder}, CodeV \cite{CodeV}, and also with the commercial LLMs without any agents.
All the experimental results can be found in Tables \ref{tab:resultsApproach1} and \ref{tab:resultsApproach2}.
It is important to note that in all implemented designs, the Supervisor and Prompt Engineer agents consistently employ the gpt-4o-mini model, while the Verilog and Testbench Generator agents utilize different LLMs.
\vspace*{-5pt}

\begin{table}[t]
    \centering
    \caption{Experimental Results using VerilogEval \cite{verilogEval}}
    \newcolumntype{?}{!{\vrule width 2pt}}
    \label{tab:resultsApproach1}
    \resizebox{\columnwidth}{!}{%
    \begin{tabular}{c?c|c|c?c|c|c} 
        \multirow{2}{*}{Model}                            & \multicolumn{3}{|c?}{VerilogEval-Machine}     & \multicolumn{3}{c}{VerilogEval-Human}         \\ \cline{2-7} 
                                                          & $pass@1$    & $pass@5$      & $pass@10$       & $pass@1$      & $pass@5$      & $pass@10$     \\ \Xhline{5\arrayrulewidth}
        GPT-4o-mini                                       & 60.1          & 62.2          & 65            & 44.2          & 49.4          & 56.4          \\ \hline
        Claude3.5-Sonnet \cite{claude3.5}                 & 74.8          & 76.9          & 79.7          & 56.4          & 60.9          & 69.9          \\ \hline
        CodeLlama-70B-Instruct \cite{codellama}           & 54.5          & 58            & 63.6          & 40.4          & 43.6          & 48.7          \\ \hline
        DeepSeek-Coder-V2-Instruct \cite{deepseekCoderV2} & 63.6          & 73.4          & 78.3          & 48.1          & 52.6          & 57.1          \\ \hline
        PyraNet-DeepSeek \cite{pyranet}                   & 77.6          & 84.6          & 89.5          & 58.3          & 62.8          & 67.9          \\ \Xhline{5\arrayrulewidth}
        RTLCoder-DeepSeek\cite{rtlcoder}                  & 61.2          & 76.5          & 81.8          & 41.6          & 50.1          & 53.4          \\ \hline
        Origen-DeepSeek \cite{origen}                     & 74.1          & 82.5          & 85.3          & 54.5          & 60.3          & 64.1          \\ \hline
        AutoVCoder-CodeQwen \cite{AutoVCoder}             & 68.5          & 79.7          & 79.7          & 48.7          & 55.8          & 55.8          \\ \hline
        CodeV-DeepSeek \cite{CodeV}                       & 77.9          & 88.6          & 90.7          & 52.7          & 62.5          & 67.3          \\ \Xhline{5\arrayrulewidth}
        VeriMind-GPT-4o-mini                              & 60.1          & \textbf{68.5} & \textbf{72.6} & 44.2          & \textbf{54.5} & \textbf{62.8} \\ \hline
        VeriMind-Claude3.5-Sonnet                         & 74.8          & \textbf{83.2} & \textbf{85.3} & 56.4          & \textbf{63.5} & \textbf{71.8} \\ \hline
        VeriMind-CodeLlama-70b-Instruct                   & 54.5          & \textbf{65  } & \textbf{70.5} & 40.4          & \textbf{48.7} & \textbf{53.8} \\ \hline
        VeriMind-Deep-Seek-Coder-V2-Instruct              & 63.6          & \textbf{76.2} & \textbf{81.8} & 48.1          & \textbf{58.3} & \textbf{65.4} \\ \hline
        VeriMind-PyraNet-DeepSeek                         & 77.6          & \textbf{90.9} & \textbf{96.5} & 58.3          & \textbf{67.9} & \textbf{74.4} \\ 
    \end{tabular}%
    }
\vspace*{-10pt}
\end{table}

\begin{table}[t]
    \centering
    \caption{Experimental Results using pass@ARC metric}
    \newcolumntype{?}{!{\vrule width 2pt}}
    \label{tab:resultsApproach2}
    \resizebox{\columnwidth}{!}{%
    \begin{tabular}{c?c|c|c?c|c|c} 
        \multirow{2}{*}{Model}                            & \multicolumn{3}{|c?}{Verilog-Machine} & \multicolumn{3}{c}{Verilog-Human} \\ \cline{2-7} 
                                                          & pass rate & ARC  & pass@ARC            & pass rate & ARC  & pass@ARC        \\ \Xhline{5\arrayrulewidth}
        GPT-4o-mini                                       & 65        & 4.48 &  57.6               & 56.4      & 5.76 & 45              \\ \hline
        Claude3.5-Sonnet \cite{claude3.5}                 & 79.7      & 3.16 &  76.1               & 69.9      & 4.7  & 60.9            \\ \hline
        CodeLlama-70B-Instruct \cite{codellama}           & 63.6      & 4.91 &  54.6               & 48.7      & 6.21 & 37.2            \\ \hline
        DeepSeek-Coder-V2-Instruct \cite{deepseekCoderV2} & 78.3      & 3.78 &  72.5               & 57.1      & 5.45 & 46.8            \\ \hline 
        PyraNet-DeepSeek \cite{pyranet}                   & 89.5      & 2.66 &  87.1               & 67.9      & 4.53 & 60              \\ \Xhline{5\arrayrulewidth}
        RTLCoder-DeepSeek\cite{rtlcoder}                  & 81.8      & 3.73 &  75.9               & 53.4      & 5.83 & 42.1            \\ \hline
        Origen-DeepSeek \cite{origen}                     & 85.3      & 2.91 &  82.3               & 64.1      & 4.81 & 55.5            \\ \hline
        AutoVCoder-CodeQwen \cite{AutoVCoder}             & 79.7      & 3.27 &  75.7               & 55.8      & 5.26 & 46.5            \\ \hline 
        CodeV-DeepSeek \cite{CodeV}                       & 90.7      & 2.45 &  89                 & 67.3      & 4.76 & 58.4            \\ \Xhline{5\arrayrulewidth}
        VeriMind-GPT-4o-mini                              & 72.6      & 4.17 &  \textbf{65.8}      & 62.8      & 5.51 & \textbf{51.3}   \\ \hline
        VeriMind-Claude3.5-Sonnet                         & 85.3      & 2.85 &  \textbf{82.5}      & 71.8      & 4.57 & \textbf{63.2}   \\ \hline
        VeriMind-CodeLlama-70b-Instruct                   & 70.5      & 4.57 &  \textbf{62.2}      & 53.8      & 5.95 & \textbf{42.1}   \\ \hline
        VeriMind-Deep-Seek-Coder-V2-Instruct              & 81.8      & 3.64 &  \textbf{76.3}      & 65.4      & 5.16 & \textbf{55  }   \\ \hline
        VeriMind-PyraNet-DeepSeek                         & 96.5      & 2.35 &  \textbf{94.8}      & 74.4      & 4.27 & \textbf{66.8}   \\ 
    \end{tabular}%
    }
\vspace*{-10pt}
\end{table}

\subsection{Pass@k Shortcoming and Solution}
One major challenge was selecting an appropriate evaluation metric.
Traditional approaches typically rely on the pass@k metric. 
However, applying pass@k to architectures that employ an iterative refinement loop can be misleading, since a single task may require multiple attempts before reaching a correct output. 
More specifically, reporting $pass@1$ for a system that iteratively corrects errors would be inaccurate if some tasks take more than one try. 
In our framework, which inherently incorporates a feedback loop to verify and fix generated code, using $pass@k$ alone would be deceptive—it could potentially report a $100\%$ $pass@1$ rate if the loop functions perfectly.

For more accurate evaluation and comparison, we adopted a dual approach. 
First, for a fair comparison, we excluded samples that required more than $k$ refinement iterations when computing pass@k for $k=1,5,10$. 
This ensures that the metric is not artificially inflated by the iterative process. 
Second, we introduced a new metric, Average Refinement Cycles (ARC), which measures the average number of iterations the system takes to produce correct Verilog code; lower ARC values indicate a more efficient process. 
This combination provides a more accurate and comprehensive evaluation of the performance of our framework.
Equation \ref{eqn:pass@arc} defines our newly proposed metric, which integrates the traditional $pass@k$ measure with the ARC to capture both the success rate and the iterative refinement efficiency.
\vspace*{-15pt}

\begin{equation} 
\label{eqn:pass@arc}
\begin{split}
     & pass@ ARC = PassRate \times e^{-0.01\times (ARC-1)^2} 
\end{split}
\vspace*{-15pt}
\end{equation}

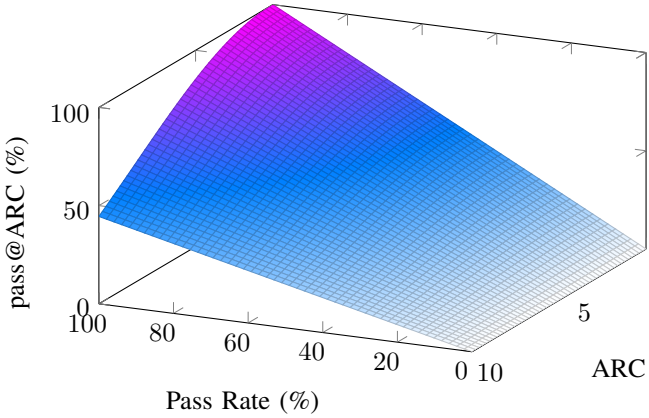
\begin{figure}[h]
\centering
\begin{tikzpicture}
  \begin{axis}[
    xlabel={Pass Rate (\%)},
    ylabel={ARC},
    zlabel={pass@ARC (\%)},
    x dir=reverse,
    domain=0:100,      
    y dir=reverse,
    y domain=1:10,     
    samples=50,
    mesh/ordering=x varies,
    colormap/cool,
    zmin=0, zmax=100,
    width=\columnwidth,
    height=0.7\columnwidth,
  ]
    \addplot3[
      surf,
    ]
    {x * exp(-0.01*(y-1)^2)};
  \end{axis}
\end{tikzpicture}
\vspace*{-10pt}
\caption{Surface plot of the pass@ARC metric: \(z = x \cdot e^{-0.01 \times (y-1)^2}\), where \(x\) is the Pass Rate and \(y\) is ARC.}
\label{fig:allValues}
\vspace*{-10pt}
\end{figure}

In this formula, PassRate represents the model’s success rate and ARC quantifies the average number of iterations required to produce correct output. 
The exponential term, $e^{-0.01\times (ARC-1)^2}$ serves as a penalty factor that equals $1$ when ARC is $1$—indicating a perfect one-shot performance—and decays as ARC increases. 
This decay penalizes models that require additional refinement iterations, with the factor $0.01$ controlling the severity of the penalty. 
By combining pass rate with ARC, the metric effectively balances accuracy with refinement efficiency, providing a comprehensive assessment of the model’s performance.
This combined metric widens the evaluative scope, allowing for a more nuanced comparison across different architectures.
It is important to note that the constant in the pass@ARC equation was chosen so that when ARC exceeds 10, the resulting metric value becomes negligibly small—effectively treating samples requiring more than 10 iterations as incomplete. 
If evaluation is needed for models that take more than 10 iterations to complete a task, this limitation can be overcome by adjusting the constant in Equation \ref{eqn:pass@arc}.
Fig. \ref{fig:allValues} presents a 3D surface plot of the proposed pass@ARC metric across its full range. 
The figure clearly demonstrates that when ARC equals 1, the metric is equivalent to pass@1, while higher ARC values lead to a gradual decrease in pass@ARC (as shown on the z-axis).
\vspace*{-5pt}

\subsection{Results Explained}
Our experiments on the VerilogEval-Machine and VerilogEval-Human benchmarks reveal that baseline models (e.g., GPT-4o-mini, Claude3.5-Sonnet, CodeLlama-70B-Instruct, DeepSeek-Coder-V2-Instruct) achieve moderate pass@1 performance, with improvements at higher pass@k levels. 
In contrast, our VeriMind models significantly boost pass@5 and pass@10 metrics. 
For example, while GPT-4o-mini attains a $60.1\%$ pass@1 on the Machine benchmark, VeriMind-GPT-4o-mini increases pass@5 and pass@10 to $68.5\%$ and $72.6\%$, respectively. 
Similarly, VeriMind-PyraNet-DeepSeek records pass@1 of $77.6\%$, pass@5 of $90.9\%$, and pass@10 of $96.5\%$. 
These results demonstrate that our feedback loop and iterative refinement process substantially enhance code generation quality without compromising initial performance.
Overall, improvements in higher-order pass@k metrics indicate that our approach increases the likelihood of generating correct Verilog code with fewer refinement cycles, as further confirmed by our pass@ARC metric.

Following the presentation of individual pass@1, pass@5, and pass@10 scores, we further evaluate our framework using the pass@ARC metric, which combines the initial pass rate with Average Refinement Cycles through an exponential penalty that increases with additional iterations. 
This design ensures that a model with high pass@1 and minimal refinement (ARC $=~$ 1) incurs little penalty, whereas additional cycles lower the pass@ARC score. 
Our experiments reveal that, while baseline models show moderate pass@k performance, the VeriMind variants consistently achieve higher pass@ARC values; for instance, VeriMind-PyraNet-DeepSeek records a pass@1 of $96.5\%$ and, with an ARC of $2.35$, a pass@ARC of $94.8\%$, significantly outperforming the baseline PyraNet-DeepSeek $(87.1\%)$. 
Similar trends are observed with the VeriMind versions of GPT-4o-mini, Claude3.5-Sonnet, CodeLlama-70B-Instruct, and DeepSeek-Coder-V2-Instruct, highlighting that our feedback loop reduces the necessary refinement iterations and provides a more comprehensive performance metric. 
Fig. \ref{fig:allPass@arc} illustrates a unified 3D visualization of the pass rate, ARC, and pass@ARC metrics.
\vspace*{-10pt}

\begin{figure}[h]
\centering
\begin{tikzpicture}
  \begin{axis}[
    title = VerilogEval-Machine ,
    view={86}{10},
    bar width = 3.5pt,
    xlabel={Pass Rate (\%)},
    ylabel={ARC},
    zlabel={pass@ARC (\%)},
    grid=both,
    width=0.97\columnwidth,
    height=0.6\columnwidth,
    xmin=60, xmax=100,
    x dir=reverse,
    xlabel style={rotate=-54, at={(axis description cs:-0.09,0.01)}, anchor=north},
    ymin=2, ymax=5,
    y dir=reverse,  
    zmin=0, zmax=100,
    legend style={at={(0.5,-0.9)}, anchor=south, draw=none, fill=none,
    font =\tiny,legend columns =2, legend cell align=left, column sep=5pt, legend image post style={color=black, solid}}]
    
    \addplot3+[
      ybar,
      fill=my_light_blue,
      draw=none,
      mark = *,
      mark options={color=my_light_blue, draw=black},
      legend image post style={color=my_light_blue, solid}] coordinates {
      (96.5,2.35,94.8) 
      };
    \addlegendentry{VeriMind-PyraNet-DeepSeek}

    \addplot3+[
      ybar,
      fill=my_dark_blue,
      draw=none,
      mark = *,
      mark options = {color=my_dark_blue, draw=black, solid},
      legend image post style={color=my_dark_blue, solid}
    ] coordinates {
      (89.5, 2.66, 87.1) 
    };
    \addlegendentry{PyraNet-DeepSeek}

    \addplot3+[
      ybar,
      fill=my_light_teal,
      draw=none,
      mark=*, 
      mark options={color=my_light_teal, draw=black},
      legend image post style={color=my_light_teal, solid}] coordinates {
      (81.8, 3.64, 76.3) 
      };
    \addlegendentry{VeriMind-DeepSeek-Coder-V2-Instruct}

    \addplot3+[
      ybar,
      fill=my_dark_teal,
      draw=none,
      mark = *,
      mark options = {color=my_dark_teal, draw=black, solid},
      legend image post style={color=my_dark_teal, solid}
    ] coordinates {
      (78.3, 3.78, 72.5) 
    };
    \addlegendentry{DeepSeek-Coder-V2-Instruct}

    \addplot3+[
      ybar,
      fill=my_light_violet,
      draw=none,
      mark=*,
      mark options={color=my_light_violet, draw=black},
      legend image post style={color=my_light_violet, solid}
    ] coordinates {
      (70.5, 4.57, 62.2) 
    };
    \addlegendentry{VeriMind-CodeLlama-70b-Instruct}

    \addplot3+[
      ybar,
      fill=my_dark_violet,
      draw=none,
      mark = *,
      mark options = {color=my_dark_violet, draw=black, solid},
      legend image post style={color=my_dark_violet, solid}
    ] coordinates {
      (63.6, 4.91, 54.6) 
    };
    \addlegendentry{CodeLlama-70B-Instruct}

    \addplot3+[
      ybar,
      fill=my_light_red,
      draw=none,
      mark=*,
      mark options={color=my_light_red, draw=black, solid},
      legend image post style={color=my_light_red, solid}
      ] coordinates {
      (85.3, 2.85, 82.5) 
    };
    \addlegendentry{VeriMind-Claude3.5-Sonnet}

        \addplot3+[
      ybar,
      fill=my_dark_red,
      draw=none,
      mark = *, 
      mark options = {color = my_dark_red, draw = black, solid},
      legend image post style={color=my_dark_red, solid}
    ] coordinates {
      (79.7, 3.16, 76.1) 
    };
    \addlegendentry{Claude3.5-Sonnet}
       
    \addplot3+[
      ybar,
      fill=my_light_green,
      draw=none,
      mark=*,
      mark options={color=my_light_green, draw=black, solid},
      legend image post style={color=my_light_green, solid}
    ] coordinates {
      (72.6, 4.17, 65.8) 
    };
    \addlegendentry{VeriMind-GPT-4o-mini}

    \addplot3+[
      ybar,
      fill=my_dark_green,
      draw=none,
      mark = *, 
      mark options = {color = my_dark_green, draw = black, solid},
      legend image post style={color=my_dark_green, solid}
    ] coordinates {
      (65  , 4.48, 57.6) 
    };
    \addlegendentry{GPT-4o-mini}

    \addplot3+[
      ybar,
      fill=my_gray,
      draw=none,
      mark=*,
      mark options={color=my_gray, solid, draw = black},
      legend image post style={color=my_gray, solid}
    ] coordinates {
      (81.8,3.73,75.9) 
    };
    \addlegendentry{RTLCoder-DeepSeek}

    \addplot3+[
      ybar,
      fill=my_yellow,
      draw=none,
      mark = *,
      mark options = {color=my_yellow, draw=black, solid},
      legend image post style={color=my_yellow, solid}
    ] coordinates {
      (85.3, 2.91, 82.3) 
    };
    \addlegendentry{Origen-DeepSeek}

    \addplot3+[
      ybar,
      fill=my_pink,
      draw=none,
      mark=*,
      mark options={color=my_pink, solid, draw = black},
      legend image post style={color=my_pink, solid}
    ] coordinates {
      (79.7, 3.27, 75.7) [CodeVCoder-CodeQwen]
    };
    \addlegendentry{AutoVCoder-CodeQwen}

    \addplot3+[
      ybar,
      fill=lime,
      draw=none,
      mark=*,
      mark options={color=lime, solid, draw = black},
      legend image post style={color=lime, solid}
    ] coordinates {
      (90.7, 2.45, 89) 
    };
    \addlegendentry{CodeV-DeepSeek}

  \end{axis}
\end{tikzpicture}
\label{fig:allPass@arc}
\vspace*{-15pt}
\vspace*{-15pt}
\end{figure}

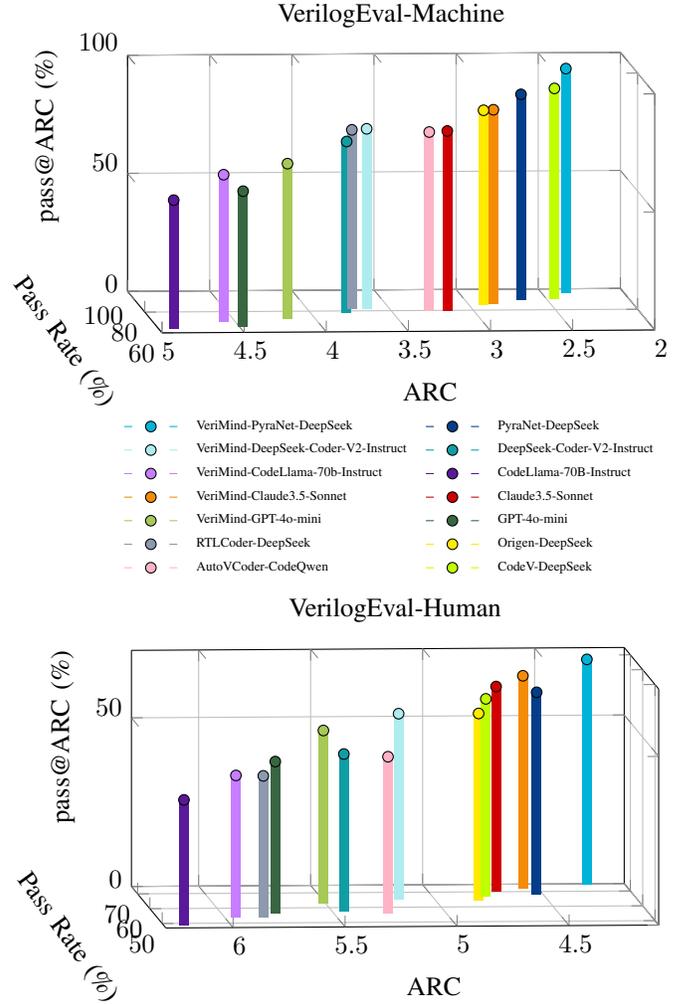
\begin{figure}[h]
\centering
\begin{tikzpicture}
  \begin{axis}[
  title = VerilogEval-Human,
    view={86}{10},
    bar width = 3.5pt,
    xlabel={Pass Rate (\%)},
    ylabel={ARC},
    zlabel={pass@ARC (\%)},
    grid=both,
    width=0.97\columnwidth,
    height=0.6\columnwidth,
    xmin=47, xmax=75,
    xlabel style={rotate=-54, at={(axis description cs:-0.09,0.01)}, anchor=north},
    x dir=reverse,
    ymin=4.1, ymax=6.3,
    y dir=reverse,  
    zmin=0, zmax=70,
    legend style={at={(0.5,1.05)}, anchor=south, draw=none, fill=none,
    font =\tiny,legend columns =2, legend cell align=left, column sep=5pt, legend image post style={color=black, solid}
}
  ]
  
    \addplot3+[
      ybar,
      fill=my_light_blue,
      draw=none,
      mark = *,
      mark options={color=my_light_blue, draw=black},
      legend image post style={color=my_light_blue, solid}] coordinates {
      (74.4, 4.27, 66.8) 
      };

    \addplot3+[
      ybar,
      fill=my_dark_blue,
      draw=none,
      mark = *,
      mark options = {color=my_dark_blue, draw=black, solid},
      legend image post style={color=my_dark_blue, solid}
    ] coordinates {
      (67.9, 4.53, 60) 
    };

    \addplot3+[
      ybar,
      fill=my_light_teal,
      draw=none,
      mark=*, 
      mark options={color=my_light_teal, draw=black},
      legend image post style={color=my_light_teal, solid}] coordinates {
      (65.4, 5.16, 55) 
      };

    \addplot3+[
      ybar,
      fill=my_dark_teal,
      draw=none,
      mark = *,
      mark options = {color=my_dark_teal, draw=black, solid},
      legend image post style={color=my_dark_teal, solid}
    ] coordinates {
      (57.1, 5.45, 46.8) 
    };

    \addplot3+[
      ybar,
      fill=my_light_violet,
      draw=none,
      mark=*,
      mark options={color=my_light_violet, draw=black},
      legend image post style={color=my_light_violet, solid}
    ] coordinates {
      (53.8, 5.95, 42.1) 
    };

    \addplot3+[
      ybar,
      fill=my_dark_violet,
      draw=none,
      mark = *,
      mark options = {color=my_dark_violet, draw=black, solid},
      legend image post style={color=my_dark_violet, solid}
    ] coordinates {
      (48.7, 6.21, 37.2) 
    };

    \addplot3+[
      ybar,
      fill=my_light_red,
      draw=none,
      mark=*,
      mark options={color=my_light_red, draw=black, solid},
      legend image post style={color=my_light_red, solid}
      ] coordinates {
      (71.8, 4.57, 63.2) 
    };

        \addplot3+[
      ybar,
      fill=my_dark_red,
      draw=none,
      mark = *, 
      mark options = {color = my_dark_red, draw = black, solid},
      legend image post style={color=my_dark_red, solid}
    ] coordinates {
      (69.9, 4.7 , 60.9) 
    };
       
    \addplot3+[
      ybar,
      fill=my_light_green,
      draw=none,
      mark=*,
      mark options={color=my_light_green, draw=black, solid},
      legend image post style={color=my_light_green, solid}
    ] coordinates {
      (62.8, 5.51, 51.3) 
    };

    \addplot3+[
      ybar,
      fill=my_dark_green,
      draw=none,
      mark = *, 
      mark options = {color = my_dark_green, draw = black, solid},
      legend image post style={color=my_dark_green, solid}
    ] coordinates {
      (56.4, 5.76, 45) 
    };

    \addplot3+[
      ybar,
      fill=my_gray,
      draw=none,
      mark=*,
      mark options={color=my_gray, solid, draw = black},
      legend image post style={color=my_gray, solid}
    ] coordinates {
      (53.4, 5.83, 42.1) 
    };

    \addplot3+[
      ybar,
      fill=my_yellow,
      draw=none,
      mark = *,
      mark options = {color=my_yellow, draw=black, solid},
      legend image post style={color=my_yellow, solid}
    ] coordinates {
      (64.1, 4.81, 55.5) 
    };

    \addplot3+[
      ybar,
      fill=my_pink,
      draw=none,
      mark=*,
      mark options={color=my_pink, solid, draw = black},
      legend image post style={color=my_pink, solid}
    ] coordinates {
      (55.8, 5.26, 46.5) 
    };

    \addplot3+[
      ybar,
      fill=lime,
      draw=none,
      mark=*,
      mark options={color=lime, solid, draw = black},
      legend image post style={color=lime, solid}
    ] coordinates {
      (67.3, 4.76, 58.4) 
    };
    
  \end{axis}
\end{tikzpicture}
\vspace*{-15pt}
\caption{Bar chart showing Pass Rate (x-axis), ARC (y-axis), and pass@ARC (z-axis) comparing the results of VeriMind and other SOTA methods.}
\label{fig:allPass@arc}
\vspace*{-10pt}
\end{figure}
\section{Conclusion and Future Works}
\label{sec:conclusionAndFutureWorks}

In this paper, we introduce VeriMind, an agentic AI framework for Verilog code generation that employs specialized agents—Supervisor, Prompt Engineer, Verilog Code Generator, Testbench Generator, and Checker—in an iterative refinement process. 
By decomposing the task into manageable subtasks and leveraging a feedback loop, VeriMind enhances both the syntactic and functional correctness of generated Verilog code while reducing refinement cycles. 
We also propose the pass@ARC metric, which combines pass@k with Average Refinement Cycles to offer a more comprehensive evaluation of initial accuracy and iterative efficiency. 
Evaluations on VerilogEval-Machine and VerilogEval-Human benchmarks show improvements of up to $8.3\%$ on pass@k and $8.1\%$ on pass@ARC, underscoring the effectiveness of our approach in advancing hardware design automation.
Future work will focus on refining the iterative process, improve security, and extend the framework to support additional hardware description languages and complex designs.

\clearpage

\clearpage
\onecolumn
\noindent \textbf{Appendix A}: \\
\label{appendix:A}
\begin{minted}[frame=lines, linenos=false, fontsize=\footnotesize]{verilog}
module traffic_light_controller(
    input wire clk,        // Clock input
    input wire reset,      // Active high reset
    output reg [2:0] north_south_lights,// For N-S dir
    output reg [2:0] east_west_lights // For E-W dir  
);
// Light encoding:
parameter RED    = 3'b100;
parameter YELLOW = 3'b010;
parameter GREEN  = 3'b001;
// State encoding
parameter S0 = 2'b00;  // NS: Green, EW: Red
parameter S1 = 2'b01;  // NS: Yellow, EW: Red
parameter S2 = 2'b10;  // NS: Red, EW: Green
parameter S3 = 2'b11;  // NS: Red, EW: Yellow
// Timing parameters 
// (adjust according to your clock frequency)
// 50M cycles = 1 second @ 50MHz
parameter GREEN_TIME  = 32'd50000000;
// 15M cycles = 0.3 seconds @ 50MHz
parameter YELLOW_TIME = 32'd15000000;  

reg [1:0] state, next_state;
reg [31:0] timer;
// State register and timer
always @(posedge clk or posedge reset) begin
    if (reset) begin
        state <= S0;
        timer <= 32'd0;
    end else begin
        if (timer == 32'd0) begin
            state <= next_state;
            // Load appropriate timer value 
            // based on next state
            case (next_state)
                S0, S2: timer <= GREEN_TIME;
                S1, S3: timer <= YELLOW_TIME;
                default: timer <= GREEN_TIME;
            endcase
        end else begin
            timer <= timer - 32'd1;
        end
    end
end
// Next state logic
always @(*) begin
    case (state)
        S0: next_state = (timer == 32'd0) ? S1 : S0;
        S1: next_state = (timer == 32'd0) ? S2 : S1;
        S2: next_state = (timer == 32'd0) ? S3 : S2;
        S3: next_state = (timer == 32'd0) ? S0 : S3;
        default: next_state = S0;
    endcase
end
// Output logic
always @(*) begin
    case (state)
        S0: begin
            north_south_lights = GREEN;
            east_west_lights = RED;
        end
        S1: begin
            north_south_lights = YELLOW;
            east_west_lights = RED;
        end
        S2: begin
            north_south_lights = RED;
            east_west_lights = GREEN;
        end
        S3: begin
            north_south_lights = RED;
            east_west_lights = YELLOW;
        end
        default: begin
            north_south_lights = RED;
            east_west_lights = RED;
        end
    endcase
end
endmodule
\end{minted}

\noindent \textbf{Appendix B}: \\
\begin{minted}[frame=lines, linenos=false, fontsize=\footnotesize]{verilog}
`timescale 1ns / 1ps

module traffic_light_controller_tb;

  // Parameters for simulation
  parameter CLK_PERIOD = 20;        // 50MHz clock (20ns period)
  parameter SIM_TIME = 5000000;     // Simulation time in ns

  // Signal declarations
  reg clk;
  reg reset;
  wire [2:0] north_south_lights;
  wire [2:0] east_west_lights;

  // Instantiate the module under test (MUT)
  traffic_light_controller UUT (
    .clk(clk),
    .reset(reset),
    .north_south_lights(north_south_lights),
    .east_west_lights(east_west_lights)
  );

  // Clock generation
  initial begin
    clk = 0;
    forever #(CLK_PERIOD/2) clk = ~clk;
  end

  // Light state definitions for easier monitoring
  wire ns_red, ns_yellow, ns_green, ew_red, ew_yellow, ew_green;
  assign ns_red    = (north_south_lights == 3'b100);
  assign ns_yellow = (north_south_lights == 3'b010);
  assign ns_green  = (north_south_lights == 3'b001);
  assign ew_red    = (east_west_lights == 3'b100);
  assign ew_yellow = (east_west_lights == 3'b010);
  assign ew_green  = (east_west_lights == 3'b001);

  // Monitor changes in light states
  initial begin
    $monitor("Time: %0t, NS: %s, EW: %s", 
             $time,
             ns_red    ? "RED   " : ns_yellow ? "YELLOW" : "GREEN ",
             ew_red    ? "RED   " : ew_yellow ? "YELLOW" : "GREEN ");
  end

  // State tracking variables
  reg [31:0] state_start_time;
  reg [31:0] state_duration;
  reg [2:0] prev_ns_lights;
  reg [2:0] prev_ew_lights;

  // Stimulus and test sequence
  initial begin
    // Initialize signals
    reset = 1;
    state_start_time = 0;
    prev_ns_lights = 3'b000;
    prev_ew_lights = 3'b000;
    
    // Apply reset
    #100 reset = 0;
    
    // Run simulation for specified time
    #SIM_TIME;
    
    // Report test results
    $display("Test completed at time %0t", $time);
    $finish;
  end

  // State transition checks
  always @(posedge clk) begin
    // Detect state changes
    if (north_south_lights != prev_ns_lights || east_west_lights != prev_ew_lights) begin
      state_duration = $time - state_start_time;
      
      // Report state duration
      if (state_start_time > 0) begin
        $display("State duration: %0t ns", state_duration);
        
        // Check timing (adjusted for clock cycles)
        if (prev_ns_lights == 3'b001 || prev_ew_lights == 3'b001) begin
          // Green state should last approximately 1 second (50M cycles * 20ns = 1000000ns)
          if (state_duration < 990000 || state_duration > 1010000) begin
            $display("WARNING:");
            $display("Green state duration (%0t ns) outside expected range", state_duration);
          end
        end
        
        if (prev_ns_lights == 3'b010 || prev_ew_lights == 3'b010) begin
          // Yellow state should last approximately 0.3 seconds (15M cycles * 20ns = 300000ns)
          if (state_duration < 290000 || state_duration > 310000) begin
            $display("WARNING:");
            $display("Yellow state duration (%0t ns) outside expected range", state_duration);
          end
        end
      end
      
      // Update state tracking
      state_start_time = $time;
      prev_ns_lights = north_south_lights;
      prev_ew_lights = east_west_lights;
    end
  end

  // Check for invalid states
  always @(posedge clk) begin
    // Check for invalid light combinations
    if (north_south_lights == 3'b000 || east_west_lights == 3'b000) begin
      $display("ERROR: Invalid light state detected at time %0t", $time);
      $display("NS lights: %b, EW lights: %b", north_south_lights, east_west_lights);
    end
    
    // Check for conflicting green signals
    if (ns_green && ew_green) begin
      $display("ERROR: Both directions have green light at time %0t!", $time);
    end
    
    // Check for conflicting yellow signals
    if (ns_yellow && ew_yellow) begin
      $display("ERROR: Both directions have yellow light at time %0t!", $time);
    end
    
    // Check that when one direction is green or yellow, the other is red
    if ((ns_green || ns_yellow) && !ew_red) begin
      $display("ERROR: East-West should be RED when North-South is GREEN or YELLOW at time %0t", $time);
    end
    
    if ((ew_green || ew_yellow) && !ns_red) begin
      $display("ERROR: North-South should be RED when East-West is GREEN or YELLOW at time %0t", $time);
    end
  end

endmodule
\end{minted}

\end{document}